\documentclass[aps,prl,twocolumn,groupedaddress]{revtex4-1}

\usepackage[dvipdfm]{graphicx}
\usepackage{dcolumn}
\usepackage{bm}

\begin{document}

\title{Common dependence on stress for the statistics of granular avalanches and earthquakes}
\author{Takahiro Hatano}
\affiliation{Earthquake Research Institute, University of Tokyo, 113-0032 Tokyo, Japan}
\author{Cl{\'e}ment Narteau}
\affiliation{Institut de Physique du Globe de Paris, Sorbonne Paris Cit\'e, Univ Paris Diderot, UMR 7154 CNRS, 1 rue Jussieu, 75238 Paris, Cedex 05, France}
\author{Peter Shebalin}
\affiliation{International Institute of Earthquake Prediction Theory and Mathematical Geophysics,Warshavskoye shosse, 79, korp 2, Moscow 113556, Russia}
\begin{abstract}
The statistical properties of avalanches in a dissipative particulate system under slow shear are
investigated using molecular dynamics simulations. It is found that the magnitude-frequency distribution
obeys the Gutenberg-Richter law only in the proximity of a critical density and that the exponent is 
sensitive to the minute changes in density. It is also found that aftershocks occur in this system with a decay
rate that follows the Modified Omori law. We show that the exponent of the magnitude-frequency
distribution and the time constant of the Modified Omori law are decreasing functions of the shear stress.
The dependences of these two parameters on shear stress coincide with recent seismological observations
[D. Schorlemmer et al. Nature {\bf 437}, 539 (2005); C. Narteau et al. Nature {\bf 462}, 642 (2009)].
\end{abstract}
\maketitle


Individual earthquakes may be regarded as large scale ruptures involving a wide range of structural
and compositional heterogeneities in the crust. However, the statistical properties of
a population of earthquakes are often described by simple power-laws.
Among them, two laws are ubiquitous and occupy a central position in statistical seismology:
the Gutenberg-Richter (GR) law \cite{Gutenberg1944} and 
the Modified Omori law (MOL) \cite{Omori1894,Utsu1970}.
The GR law describes the earthquake magnitude-frequency distribution
\begin{equation}
\label{GR}
P (M) \propto 10^{-b M_w}, 
\end{equation}
where $M_w$ is the moment magnitude and $b$ a constant with a value around $1$ along active fault zones.
The MOL describes the aftershock occurrence rate
\begin{equation}
\label{Omori}
n(t) \propto (t + c)^{-p}, 
\end{equation}
where $t$ is the elapsed time from the triggering event (the so-called mainshock), 
$p$ a positive non-dimensional constant with a typical value of $1$,  
and $c$ is a time constant.
The parameters in these two laws are believed to bear some information on the physical state of the crust.
Indeed, Schorlemmer et al. find that the $b$-value of the GR law is decreasing going from normal
(extension) over strike-slip (shear) to thrust (compression) earthquakes \cite{Schorlemmer2005}. 
Narteau et al. also find that the time constant $c$ in the MOL has the same dependence 
on the faulting mechanism \cite{Narteau2009}.
These two observations indicate that, under a simple assumption, $b$ and $c$ are decreasing functions
of shear stress. Although the underlying mechanism needs further investigation, this may reflect a common
time-dependent behavior of fracturing in rocks during the propagation of earthquake ruptures and the
nucleation of aftershocks.

Because the shear stress along an active fault is not directly measurable, a solution to 
address stress dependences in earthquake statistics is to analyze models that
implement a restricted set of physical processes.
There are indeed many models that resemble seismicity, ranging from rock fracture
experiments \cite{Mogi1962,Scholz1968,Hirata1987} to computer simulations on cellular
automata \cite{Olami1992}.
Among them, sheared granular media
\cite{Dalton2001,Dalton2002,Twardos2005,Bretz2006,Pena2009}
fit our purpose perfectly because it is a simple representation of a granular fault gouge for which 
both the energy and the stress can be easily defined.
Here, we perform numerical simulations showing that, as for real seismicity,
avalanches in sheared granular matter obeys the GR law and the MOL with $b$ and $c$-values 
which are decreasing functions of the shear stress.

Our granular system is made of frictionless spheres with diameters
of $d$ and $0.7d$ (the ratio of populations is $1:1$).
For the sake of simplicity, we assume that the mass $M$ of these particles are the same.
We limit ourselves to a rather small-size system ($N=1500$) for computational efficiency.
Using the radius and the position of particle $i$, which are denoted by $R_i$ and 
${\bf r}_{i}$, respectively, the force between particles $i$ and $j$ is written as 
${\bf F}_{ij}= \left[kh_{ij} - \zeta{\bf n}_{ij}\cdot\dot{{\bf r}}_{ij}\right]{\bf n}_{ij}$.
Here ${\bf r}_{ij}={\bf r}_{i}-{\bf r}_{j}$, ${\bf n}_{ij}={\bf r}_{ij}/|{\bf r}_{ij}|$, and $h_{ij} = (R_i+R_j) -|{\bf r}_{ij}|$ is the overlap length.
If $R_i+R_j < |{\bf r}_{ij}|$, particles $i$ and $j$ are not in contact so that the force vanishes.
Throughout this study, we adopt the units in which $d=1$, $M=1$, and $k=1$.
We choose $\zeta = 2.0$, which corresponds to the vanishing coefficient of restitution.
A constant shear rate $\dot\gamma$ is applied to the system through the Lees-Edwards boundary conditions \cite{Allen}.
Note that under these boundary conditions the system volume is constant.
Thus, the important parameters are the shear rate $\dot\gamma$ and the packing fraction $\phi$.
A steady state of uniform shear rate $\dot\gamma$ can be realized starting from a class of special initial conditions.
Here, we investigate such uniform steady states.

As observed in many previous works on amorphous systems \cite{Pena2009,Bretz2006,Twardos2005,Durian1997,Tewari1999,Kabla2007,Maloney2004,Lerner2010,Heussinger2010},
the temporal fluctuations of the energy becomes volatile if the shear rate is sufficiently low and the density is sufficiently high.
From the temporal fluctuation of the energy $E(t)$, we define an energy drop event as follows.
The beginning of an event is defined as the time $t=t_1$ at which the energy $E(t)$ starts decreasing; i.$\;$e.\ , $\dot E(t_1)=0$ and $\ddot E(t_1) < 0$.
The definition of the end of an event is the opposite; 
$t=t_2$ at which $\dot E(t_2)=0$ and $\ddot E(t_2) > 0$.
Such an energy drop event is referred to as an {\it avalanche} throughout this paper.
We also calculate the (global) shear stress ${\hat\sigma}(t)$ using the virial \cite{Zubarev}.
Important quantities that characterize an avalanche are the magnitude 
$M\equiv \log_{10}[E(t_1)-E(t_2)] + 11$, the initial stress $\sigma\equiv{\hat\sigma} (t_1)$, 
the stress drop $\Delta\sigma\equiv{\hat\sigma}(t_1) - {\hat\sigma}(t_2)$, 
and the duration $T\equiv t_2-t_1$.
Note that, with this definition of magnitude, the GR law reads $P(M)\propto 10^{-2/3 bM}$,
as $M_w \simeq 2/3 \log_{10}[E(t_1)-E(t_2)]+{\rm const}.$
Hereafter, we use $\beta\equiv 2/3b$ instead of $b$.
Note also that here the spatial information of avalanches is discarded.
In this case, one might overlook simultaneous avalanches occurring at different places.
However, this is unlikely because the present system is small (i.$\;$e.\ the characteristic length 
is approximately of $9d$).

First, we discuss the nature of the avalanche magnitude-frequency distribution.
\begin{figure}
\includegraphics[width=9cm]{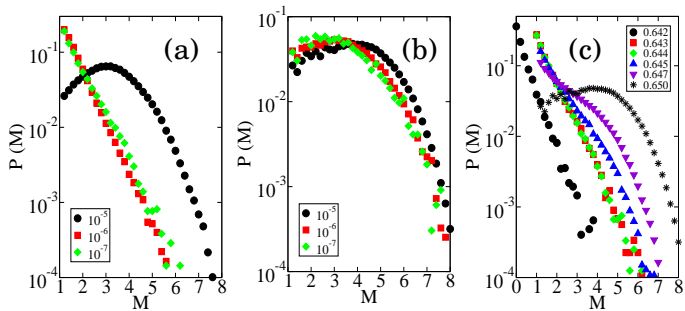}
\caption{\label{fig1} Avalanche magnitude-frequency distributions for several densities and shear rates.
(a) $\phi=0.644$. (b)  $\phi=0.650$. (c) $\dot\gamma=10^{-7}$ except for $\phi=\{0.647,\, 0.650\}$ 
($\dot\gamma=10^{-8}$).}
\end{figure}
Figures \ref{fig1} (a) and (b) show these distributions at several shear rates for 
$\phi=0.644$ and $\phi=0.650$, respectively.
The distribution is independent of the shear rate at sufficiently low shear rates.
The characteristic shear rate below which the distribution is rate-independent
depends on the density and may be interpreted as the inverse of the structural relaxation time.
In the density range investigated here, the characteristic shear rate is a decreasing function of the density.
For example, it is approximately $10^{-6}$ for $\phi=0.644$ and $10^{-8}$ for $\phi=0.650$.
Hereafter, we discuss such rate-independent behaviors by choosing sufficiently low shear rates.

Figure \ref{fig1} (c) shows rate-independent magnitude-frequency distributions at several densities.
They may be regarded as the GR law in the proximity of a critical density ($\phi\simeq 0.644$),
whereas they no longer obey the GR law at higher densities. At much lower densities
($\phi\le0.642$), an avalanche is no longer observed. It should be noted that the $\beta$-value 
seen in the proximity of a critical density is sensitive to the minute changes in density.
The $\beta$-value is a decreasing function of the density ranging from $0.47$ 
to $0.78$. We obtain the smallest value ($0.47$) at the largest density ($\phi=0.645$)
and the largest value ($0.78$) at the smallest density ($\phi=0.642$).
The $\beta$-value at intermediate densities ($\phi=\{0.643,\, 0.644\}$) is approximately $0.64$.
This range of values is comparable to that of earthquakes for which the $\beta$-value
varies from $0.50$ along normal faults (i.$\;$e.\ , extensional regime, low stress) 
to $0.73$ along reverse faults (i.$\;$e.\ , compressional regime, high stress).

We remark that distribution functions of other quantities also obey power laws.
Among them, we find that the distribution functions of the duration $T$ exhibit
power-law tails, the exponent of which is approximately $3.0$ irrespective of the density.
We also find that the exponent for the stress drop distribution is twice larger than that for energy drop.

As similar power law behaviors have been observed in some granular systems, it is interesting
to compare the present result with other studies. A major interest is the $\beta$-value: 
$0.36$ to $0.95$ \cite{Dalton2002}, $0.82$ to $0.89$ \cite{Pena2009}, and $1$ \cite{Bretz2006} 
have been reported, whereas the GR law is not observed in an experiment \cite{Twardos2005}.
Although such non-universality may be perplexing to statistical physicists, it may be rather rational 
in view of the present study; namely, the GR law holds only in the proximity of a critical density 
and the $\beta$-value depends on both the density and the stress level.
Thus, the $\beta$-value may not be an analogue for critical exponents.
As an evidence of such non-universality, one can further illustrate a wide range of $\beta$-values 
obtained in various amorphous systems \cite{Durian1997,Tewari1999,Kabla2007,Maloney2004,Lerner2010,Heussinger2010}.
Thus, there may exist many unknown ingredients that control the $\beta$-value, 
and further investigation is required for the unified understanding of all these exponents
in general amorphous systems.

To discuss the effect of shear stress on the $\beta$-value, we introduce the shear stress $\sigma$
at the beginning of an event as an additional argument to the magnitude-frequency distribution. Then,
$P(M, \sigma)$ is the conditional probability of observing a magnitude $M$ avalanche under
the (global) shear stress value $\sigma$. For convenience, $\sigma$ is integrated in a certain
interval $S_i \equiv [10^{-i/2}, 10^{-i/2+0.5}]$ $(i=\{1,\, 2,\, \cdots,\, 9\})$. Thus, we obtain the following
distribution function: $P_i (M) = \int_{S_i} {\rm d}\sigma P(M, \sigma)$. 
\begin{figure}
\includegraphics[width=8cm]{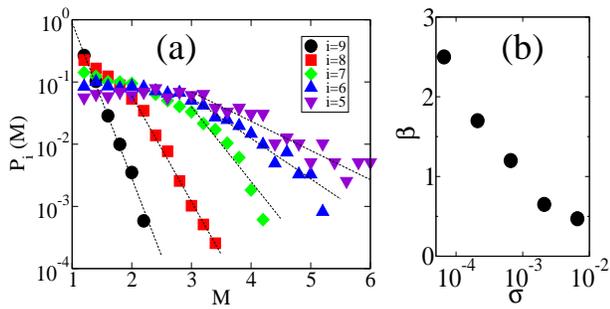}
\caption{\label{fig2} Avalanche magnitude-frequency distribution for different
ranges of shear stress values (see text for the definition).
The $\beta$-value decreases at higher shear stress.
The volume fraction is $\phi=0.643$ and the shear rate is $\dot\gamma=10^{-7}$.
}
\end{figure}
Figure \ref{fig2} shows the behaviors of $P_i(M)$ at $\phi=0.643$. We can see that the
probability of observing a larger avalanche increases as the shear stress increases. More
importantly, the distribution function at each stress level develops the tail, with
the $\beta$-value being a decreasing function of the shear stress. We confirm that this stress
dependence is independent of the density in the range $0.642\le\phi\le 0.645$. In addition,
the shear stress dependence of the $\beta$-value is qualitatively the same as that 
in rock fracture experiments \cite{Scholz1968}.

Because aftershocks result from changes of stress induced by a mainshock, we disregard 
the smaller avalanches for which $M<0$ and consider ranges of magnitude for mainshocks                                              
$[M^{\rm M}_{\rm min}, M^{\rm M}_{\rm max}]$
and aftershocks 
$[M^{\rm A}_{\rm min}, M^{\rm A}_{\rm max}]$.
In practice, a magnitude $M\in[M^{\rm M}_{\rm min}, M^{\rm M}_{\rm max}]$ event 
occuring at time $t_{\rm M}$ is not
a mainshock if there is at least one avalanche of the same or higher magnitude range in the
time interval $[t_{\rm M}-\Delta T, t_{\rm M}+\Delta T]$. Thus, we only consider isolated
mainshocks and, taking sufficiently large $\Delta T$-values, we also avoid overlapping aftershock
sequences that belong to different mainshocks. Then, all $M<M^{\rm M}_{\rm min}$ avalanches
that follow a magnitude $M^{\rm M}$ mainshock within the time interval $[t_M, t_M+\Delta T]$
are regarded as aftershocks. Finally, each aftershock is characterized by its magnitude $M^{\rm A}$
and the elapsed time $\tau$ since the mainshock.

In order to reduce artifacts related to event detectability, we disregard larger magnitude
range for mainshocks and smaller magnitude range for aftershocks. In addition, the magnitude
range of mainshocks are chosen within the intermediate magnitude range in which the GR law holds.
Using this strict methodology, we considerably reduce the number of events in our artificial
catalogues. Therefore, in all ranges of magnitudes, aftershocks are stacked with respect to the
time of their mainshocks to finally end-up with a single aftershock sequence.

\begin{figure}
\includegraphics[width=9cm]{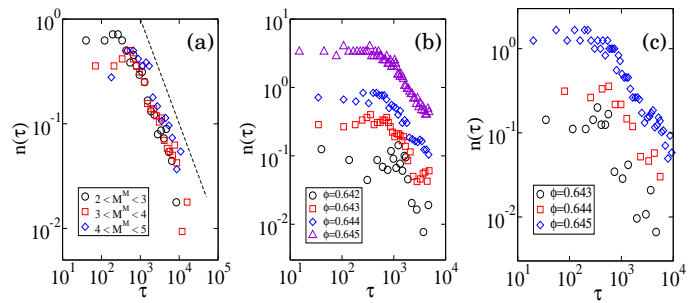}
\caption{\label{Omoris} Aftershock frequencies as functions of the elapsed time from the mainshock.
The shear rate is $\dot\gamma=10^{-7}$.
Aftershocks are stacked for several ($\ge 10$) mainshocks and the number of aftershocks for each data set is typically larger than $1000$.
(a) $\phi=0.644$, $M^{\rm A} \in [1, 3]$, $\Delta T=2\times10^{4}$. The dashed line is proportional to $1/t$.
(b) $M^{\rm M}\in[3, 4]$, $M^{\rm A}\in[1, 3]$, $\Delta T=5\times10^{3}$.
(c) $M^{\rm M}\in[4, 5]$, $M^{\rm A}\in[2, 4]$, $\Delta T=1\times10^{4}$.}
\end{figure}
Figure \ref{Omoris} (a) shows the aftershock frequencies for several choices of 
$[M^{\rm M}_{\rm min}, M^{\rm M}_{\rm max}]$. The magnitude range of aftershocks is chosen as $[1, 3]$.
We find that the MOL, Eq. (\ref{Omori}), holds with the exponent $p\simeq 1$ irrespective of the magnitude range of mainshocks.
Figures \ref{Omoris} (b) and (c) show the density dependence of aftershock frequencies.
We test several densities and shear rate ($\phi=\{0.642,\, 0.643,\, 0.644,\, 0.645\}$ and $\dot\gamma=\{10^{-6},\, 10^{-7},\, 10^{-8}\}$)
to find that the exponent $p\simeq 1$ is insensitive to the values of these parameters. This robustness makes a contrast to the
sensitivity of the exponent $\beta$ in the GR law to the density.
We also find that the time constant $c$ is independent of the density.
More importantly, we find that the time constant $c$ is independent of the shear rate. This indicates that aftershocks are not
driven by the additional shear strain applied after a mainshock; rather, they are caused by the intrinsic relaxation dynamics.
Thus, we do not expect aftershocks in the quasi-static shear deformation, in which a system fully relaxes to a stable
configuration after an avalanche.

\begin{figure}
\includegraphics[width=6cm]{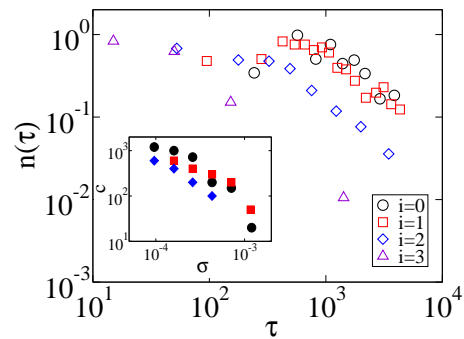}
\caption{\label{fig4} Aftershock frequencies at $\phi=0.644$ for several stress ranges; 
$\sigma=[\exp(i-10), \exp(i-9)]$ $(i=0, 1, 2, 3)$.
The magnitude ranges are $M^{\rm M}\in[3, 4]$, and $M^{\rm A}\in[1, 3]$.
Each frequency is normalized as $n(0)=1$ so that the stress dependence of $c$ is apparent.
The inset shows the stress dependence of the $c$-value.
Circles: $\phi=0.644$, $M^{\rm M}\in[3, 4]$, and $M^{\rm A}\in[1, 3]$.
Squares: $\phi=0.645$, $M^{\rm M}\in[3, 4]$, and $M^{\rm A}\in[1, 3]$.
Diamonds: $\phi=0.644$, $M^{\rm M}\in[2, 3]$, and $M^{\rm A}\in[1, 2]$.}
\end{figure}
Next, we show that the time constant $c$ depends on the shear stress.
We define for aftershocks the stress range $[\sigma_{\rm min}, \sigma_{\rm max}]$ and select only aftershocks that
belong to this stress range and the magnitude range $[M^{\rm A}_{\rm min}, M^A_{\rm max}]$. The aftershock
frequencies are shown in Figure \ref{fig4}, in which the MOL (with $p=1$) holds clearly and $c$ is a decreasing
function of shear stress. We estimate the $c$-values by fitting the data with $A/(\tau+c)$ using the
maximum-likelihood method. As shown in the inset of Figure \ref{fig4}, the $c$-value has a negative dependence
on the shear stress.
We confirm this stress dependence for two densities ($\phi=\{0.644,\, 0.645\}$) and for two magnitude ranges
($M^{\rm M}\in[3, 4]$, $M^{\rm A}\in[1, 3]$ and $M^{\rm M}\in[4, 5]$, $M^{\rm A}\in[2, 4]$). This negative
shear-stress dependence of the $c$-value for aftershocks is similar to those suggested by seismological observations
and various models of aftershock production. As in models based on friction, damage mechanics, or static
fatigue, our numerical experiment shows that a power-law relaxation regime may be delayed for an amount of
time which is inversely proportional to the global shear stress.

To conclude, two fundamental laws in earthquake statistics (GR and MOL) are also relevant to avalanches in
sheared granular matter. Most importantly, the stress dependences of the parameters in these laws are 
the same as those observed in real seismicity. This coincidence may be rather remarkable in view of the 
simpleness of the simulation model and the complexity of the crust.
A quantitative result in the present study, that the time constant $c$ in MOL is inversely proportional to
the applied shear stress, may provide seismologists with a new tool for inferring the stress level in the crust.
It may also be relevant to prevention of silo avalanches, because one can infer the stress level in silos by
observing a time series of acoustic emissions. To address such applications, the present result must be tested 
in more realistic and complex systems that are relevant to seismogenic zones and various industrial situations.

The author (TH) acknowledges helpful discussions with H. Matsukawa and M. Otsuki.

\end{document}